\documentclass[tightenlines,superbib,floats,rmp,11pt]{revtex4}
\usepackage{amssymb}
\usepackage{epsfig}
\usepackage{amsmath}

\begin{document}

\title{Complex Networks: Time--Dependent Connections and Silent Nodes}
\author{J. Marro$^{\ast }$, J.J. Torres$^{\ast }$, and J.M. Cortes$^{\ast \S %
}$}
\affiliation{$^{\ast }$Institute \textit{Carlos I}\ for Theoretical and Computational
Physics, and \textit{Departamento de Electromagnetismo y F\'{\i}sica de la
Materia}, University of Granada, Facultad de Ciencias, 18071 Granada, Spain\\
$^{\S }$Institute for Adaptive and Neural Computation, University of
Edinburgh, EH1 2QL, UK.}

\begin{abstract}
We studied, both analytically and numerically, complex excitable networks,
in which connections are time dependent and some of the nodes remain silent
at each time step. More specifically, (\textit{a}) there is a heterogenous
distribution of connection weights and, depending on the current degree of
order, some connections are reinforced/weakened with strength $\Phi $ on
short--time scales, and (\textit{b}) only a fraction $\rho $ of nodes are
simultaneously active. The resulting dynamics has attractors which, for a
range of $\Phi $ values and $\rho $ exceeding a threshold, become unstable,
the instability depending critically on the value of $\rho .$ We observe
that (\textit{i}) the activity describes a trajectory in which the close
neighborhood of some of the attractors is constantly visited, (\textit{ii})
the number of attractors visited increases with $\rho ,$ and (\textit{iii})
the trajectory may change from regular to chaotic and vice versa as $\rho $
is, even slightly modified. Furthermore, (\textit{iv}) time series show a
power--law spectra under conditions in which the attractors' space is most
efficiently explored. We argue on the possible qualitative relevance of this
phenomenology to networks in several natural contexts.\smallskip 
\end{abstract}

\maketitle

\section{Introduction and motivation}

The concept of a \textit{network} ---defined as a sufficiently large set of
nodes connected in pairs by edges--- is potentially useful to help our
understanding of the cooperative phenomena which are behind complex behavior
in science and technology. Therefore, there has been a great interest for
networks in physics during the last decade or so \cite{nets1,nets2,nets3}.
Most of these studies have focused on the case in which the edge between any
two nodes is either present or not. This relatively simple situation allows
one to investigate, in particular, \textit{wiring} topology which, for
example, has lead to the discovery of \textit{scale--free} and \textit{%
small--world} networks in natural and man--made systems. However%
, real networks exhibit a number of relevant qualities besides interesting
topological structure \cite%
{nets3,complex1,complex2,complex2a,complex3,complex4}. In this paper, we are
concerned with two features which could be essential to a network demeanor:

\noindent\ (\textit{a}) \textit{Weighted and time--dependent connections.}
Very generally, intensities and/or capacities vary notably from one edge to
the other in actual networks. For instance, a main feature of trophic webs
is the complexity of pattern flows along the food chains, the agents in
social and communication (e.g., cell phone) networks exchange assets or
information according to various rules and depending on their partners,
transport connections differ in capacity and actual number of transits and
passengers, and effective ionic interactions constantly vary in 
condensed matter due to reactions as well as to diffusion and local
rearrangements of ions and impurities. It is to be stressed that the
connection weights in these cases often vary with time. There are variations
of weights on a long--time scale. Their main purpose seems to be determining
the nature and degree of heterogeneity the network needs for its intended
function, say, computation, transport, cooperation, etc. In addition,
although perhaps less investigated yet, weights may change on a short--time
scale to improve actual functioning. To our knowledge, the best documented
cases so far of such \textit{fast fluctuations }do not belong to physics but
to computational and neural networks. As a matter of fact, the human brain
is the paradigm of a weighted network \cite{brainweight,9bis}, and it is
also clear--cut that high--level functions in the brain rely on fast
synaptic changes during operation \cite{abb}. Consequently, as we have a
main interest here on short--time weight variations, we shall in the
following often use the language and refer to observations on neural and,
eventually, computational networks. In any case, our setting is rather
general and we believe that the main behavior described in this paper should
apply to networks in different contexts (see, for instance, Refs. \cite%
{complex2,complex3}).

\noindent\ (\textit{b}) \textit{Partial activation of nodes.} In addition to
the above ---and perhaps also a further source of \textit{fast fluctuations}%
---, one may argue that there is no need for a network to maintain \textit{%
all} the nodes fully informed of the activity of \textit{all} the others at 
\textit{all} times. Relaxing such situation would both simplify the case and
turn operation more economical. Moreover, there are some indications that
certain nodes are more active than others, and that only a fraction of nodes
is actually engaged at each time in some cooperative tasks. For example,
this is the characteristic behavior of excitable media in which elements,
after responding to perturbation, are refractory to further excitation \cite%
{exci,julyan}. This is interesting because such behavior could sometimes
reveal to the observer as (relatively) fast time variations of connections
as described above. The possibility of having reticent nodes is also a
recent concern in computer science in relation with parallelism \cite%
{computer0,computer}, in mathematical--physics \cite{phys}, and in
neuroscience, where it has been associated with working memories \cite%
{micro,workmem,wage}, variability of neuron thresholds \cite{threshold} and
silent neurons \cite{silent,silent2}. On the other hand, there is evidence
of partial synchronization in many different situations \cite{sincro}. In
principle, this is a different phenomenon but one may argue that some of the
observed partial synchronization processes, in which some elements do not
attend to the others' mode, could be associated with the existence of silent
and/or excitable units, the case of interest here. In any case, studying the
effect of updating only a fraction of nodes will certainly shed light on the
possible consequences of having partial synchronization in the network.

The investigation of (\textit{a}) in physics has only recently been
initiated. As an example we mention the observed aging of nodes, e.g., in
the networks of scientific publications where original papers stop receiving
links after a finite time because review papers are then cited instead \cite%
{22bis}; see also, for instance, Refs.\cite%
{complex1,complex2,complex2a,complex3,complex4}. However, studying the
consequences of fast connection changes in biologically inspired models has
already a two--decades history ---see \cite{cortesNC} and references
therein. For example, it has recently been shown that the susceptibility of
a network to outside influence increases dramatically for excitable nodes 
\cite{mauro} and, more specifically, under a competition of processes which
tend to increase and decrease, respectively, the efficiency of synaptic
connections at short times \cite{facilitation}. To the best of our
knowledge, investigation of feature (\textit{b}) is rarer \cite%
{herzPRE,phys,computer,LinksNodes}, in spite of the fact that there is some
---as mentioned above--- specific motivation for it in several fields.
Trying to understand the combined effect of these two features, (\textit{a})
and (\textit{b}), is a main objective here. We show that varying the
fraction of nodes that are simultaneously active induces a variety of
qualitatively different behaviors when the system is in a state of great
susceptibility, but not in more general conditions. The susceptibility
needed to observe the most interesting behavior is shown to occur under
appropriate tuning of the connection weights with the network activity. It
thus ensues that the effects of (\textit{a}) and (\textit{b}) are correlated
with each other ---which confirms a suspicion mentioned in the description
of (\textit{b}) above. Even more, it seems that the concurrence of (\textit{a%
}) and (\textit{b}) could be needed in some occasions in nature. As a first
application, we describe here how a model exhibits unstable dynamics, which
leads to itinerancy and chaotic behavior in a way that mimics both general
expectations and some recent biological observations.

\section{Definition of a simple dynamic network model}

A full description of the network configuration requires both the set of
node states or \textit{activities}, \textbf{$s$}$\equiv \left\{
s_{i}\right\} ,$ and the set of connection weights, $\mathbf{w\equiv }%
\left\{ w_{ij}\in 
\mathbb{R}
\right\} ,$ where, $i,j=1,\ldots ,N.$ From these we define a local field on
each node due to the weighted action of the others, namely, $h_{i}\left( 
\mathbf{s},\mathbf{w}\right) \equiv \sum_{j\neq i}w_{ij}s_{j}.$ At each time
unit, one updates the activity of $n$ nodes, with $1\leqslant n\leqslant N.$
This induces evolution in discrete time, $t,$ of the configuration
probability distribution according to the master equation:%
\begin{equation}
P_{t+1}(\mathbf{s})=\sum_{\mathbf{s}^{^{\prime }}}\text{\textbf{T}}\left( 
\mathbf{s}^{\prime }\mathbf{\rightarrow s}\right) P_{t}(\mathbf{s}^{\prime
}),  \label{Meq}
\end{equation}%
where the transition probability may be written as 
\begin{equation}
\text{\textbf{T}}\left( \mathbf{s\rightarrow s}^{\prime }\right) =\sum_{%
\mathbf{x}}p_{n}(\mathbf{x})\prod_{\left\{ i|x_{i}=1\right\} }\tau
_{n}\left( s_{i}\rightarrow s_{i}^{\prime }\right) \prod_{\left\{
i|x_{i}=0\right\} }\delta _{s_{i},s_{i}^{\prime }}.  \label{rate}
\end{equation}%
Here, $\mathbf{x}$ is an operational set of binary indexes ---fixed to $1$
at $n$ sites chosen at each time according to distribution $p_{n}\left( 
\mathbf{x}\right) ,$ and fixed to zero at the other $N-n$ sites. The choice (%
\ref{rate}) simply states that one (only) updates simultaneously the
selected $n$ nodes. The corresponding elementary rate is%
\begin{equation}
\tau _{n}\left( s_{i}\rightarrow s_{i}^{\prime }\right) =\sigma \left(
s_{i}\rightarrow s_{i}^{\prime }\right) \left[ 1+\left( \delta
_{s_{i}^{\prime },-s_{i}}-1\right) \delta _{n,1}\right] ,  \label{3}
\end{equation}%
where $\sigma =\sigma \left( \mathbf{s},\beta \right) $ is a function to be
determined, with $\beta $ an inverse temperature parameter.

The above describes \textit{parallel updating}, as in cellular automata,%
\textit{\ }for $n=N$ or, macroscopically, $\rho \equiv n/N\rightarrow 1.$
However, the model describes \textit{sequential updating}, as in kinetic
magnetic models, for $n=1$ or $\rho \rightarrow 0.$ We are interested in
changes with $\rho \in \left( 0,1\right) .$ In addition to allow for a
sensible generalization of familiar cellular automata, this bears some
practical interest, as indicated in the introduction. For example, assuming
a neural network, $\rho $ may stand for the fraction of neurons that are
stimulated each cycle. There is no input on the other $1-\rho ,$ so that
information from the previous state is maintained. This induces persistent
activity which has been argued to be a basis for working memory \cite%
{micro,workmem,wage}. Varying $\rho $ may also be relevant to simulate the
observed variability of the neurons' threshold \cite{threshold} and the
possible existence of \textit{silent neurons} \cite{silent} or \textit{dark
neuro--matter }\cite{silent2}, for instance. These are just examples of the
fact that varying $\rho $ has a great general interest to better
understanding excitable media.

The equations above may be simulated in a computer for different choices of $%
p_{n}$ and transition details. In order to obtain analytical results,
however, we need to simplify the model somewhat. That is, we shall refer to
the case in which the node activities are binary, $s_{i}=\pm 1,$ the $n$
nodes to be updated are chosen at random, so that one has $p_{n}\left( 
\mathbf{x}\right) =\tbinom{N}{n}^{-1}\delta \left( \sum_{i}x_{i}-n\right) ,$
and $\sigma $ in (\ref{3}) is an arbitrary function of (only) $\beta
s_{i}h_{i}$ which satisfies detailed balance. In spite of the latter,
detailed balance is not fulfilled by the superposition \textbf{T} for $n>1,$
so that resulting steady states are generally out of equilibrium, which is
known to be realistic \cite{marroB}. On the other hand, for simplicity, in
order to be able to obtain some analytical results, we shall assume that the
fields are such that $h_{i}\left( \mathbf{s},\mathbf{w}\right) =h\left[ 
\mathbf{\pi }\left( \mathbf{s}\right) ,\mathbf{\xi }_{i}\right] .$ Here, $%
\mathbf{\xi }_{i}\equiv \left\{ \xi _{i}^{\mu }=\pm 1;\mu =1,\ldots
,M\right\} $ stands for $M$ given realizations of the set of activities, or 
\textit{patterns,} and $\mathbf{\pi }\equiv \left\{ \pi ^{\mu }\left( 
\mathbf{s}\right) \right\} ,$ where 
\begin{equation}
\pi ^{\mu }\left( \mathbf{s}\right) =N^{-1}\sum_{i}\xi _{i}^{\mu }s_{i},
\end{equation}%
measures the \textit{overlap} between the current state and pattern $\mu .$
For $N\rightarrow \infty $ and finite $M,$ i.e., in the limit $\alpha \equiv
M/N\rightarrow 0,$ the time equation%
\begin{equation}
\pi _{t+1}^{\mu }\left( \mathbf{s}\right) =\rho N^{-1}\sum\nolimits_{i}\xi
_{i}^{\mu }\tanh \left\{ \beta h_{i}\left[ \mathbf{\pi }_{t}\left( \mathbf{s}%
\right) ,\mathbf{\xi }_{i}\right] \right\} +\left( 1-\rho \right) \pi
_{t}^{\mu }\left( \mathbf{s}\right)  \label{mt}
\end{equation}%
follows for any $\mu .$ Actual applications concern finite values for both $%
M $ and $N,$ so that the limit $\alpha \rightarrow 0$ is not very
intesreting in practice. This and other restrictions are not essential to
the model, however; in fact, our simulations below concern more general
situations, as pointed out when necessary.

\begin{figure}[ht!]
\centerline{\psfig{file=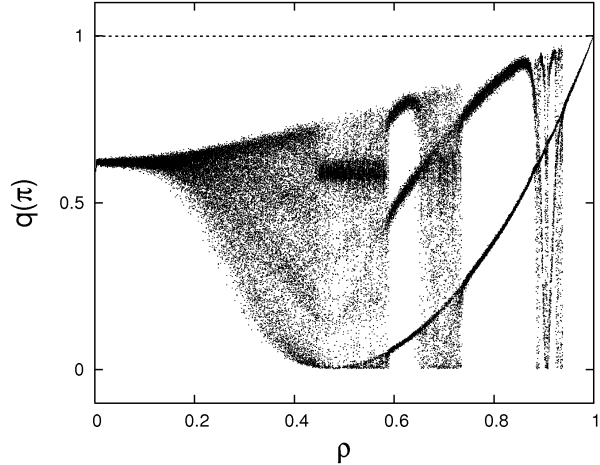,width=8.5cm}}
\caption{Evidence of chaos. Bifurcation diagram showing the stationary order parameter $q\left( \mathbf{\protect\pi }%
\right) ,$ as defined in the main text, \textit{versus the synchronization }%
parameter $\protect\rho $ for $M=20$ random patterns, $N=3600$ nodes, $%
\protect\beta =20$ and $\Phi =-1/2.$ This behavior is characteristic of any $%
\Phi \neq 1,$ and it follows indistinctly from the analytical solution and
from Monte Carlo simulations.}
\label{figure1n}
\end{figure}

The model allows for different relations between the fields $h_{i}$ and the
other network properties. The simplest case at hand for specific relations
of such kind is Hopfield's \cite{hopf1} which follows here for $\rho
\rightarrow 0$ and weights fixed according to the Hebb prescription, i.e., $%
w_{ij}=N^{-1}\sum_{\mu }\xi _{i}^{\mu }\xi _{j}^{\mu }.$ The symmetry $%
w_{ij}=w_{ji}$ then assures $P_{t\rightarrow \infty }\left( \mathbf{s}%
\right) \propto \exp \left( \beta \sum_{i}h_{i}s_{i}\right) .$ This
corresponds to thermodynamic equilibrium and ---using the neural--network
argot--- this is a case that exhibits \textit{associative memory}. This
means that, for high enough $\beta ,$ the patterns $\left\{ \mathbf{\xi }%
_{i}\right\} $ are attractors of dynamics \cite{hopf2}, as if they would
have been \textit{stored} in the connections and recalled in the course of
the system relaxation with time. Equilibrium is generally impeded for $\rho
>0$ \cite{grin}, and the asymptotic state then strongly depends on dynamic
details \cite{marroB,odorRMP}. We checked that, in agreement with some
indications \cite{herzPRE}, the Hopfield--Hebb network also exhibits
associative memory for $\rho >0.$ However, no new physics emerges as $\rho $
is varied in this case, and it is likely this occurs rather generally
concerning dynamics for simple weighted networks.
\begin{figure}[ht!]
\centerline{\psfig{file=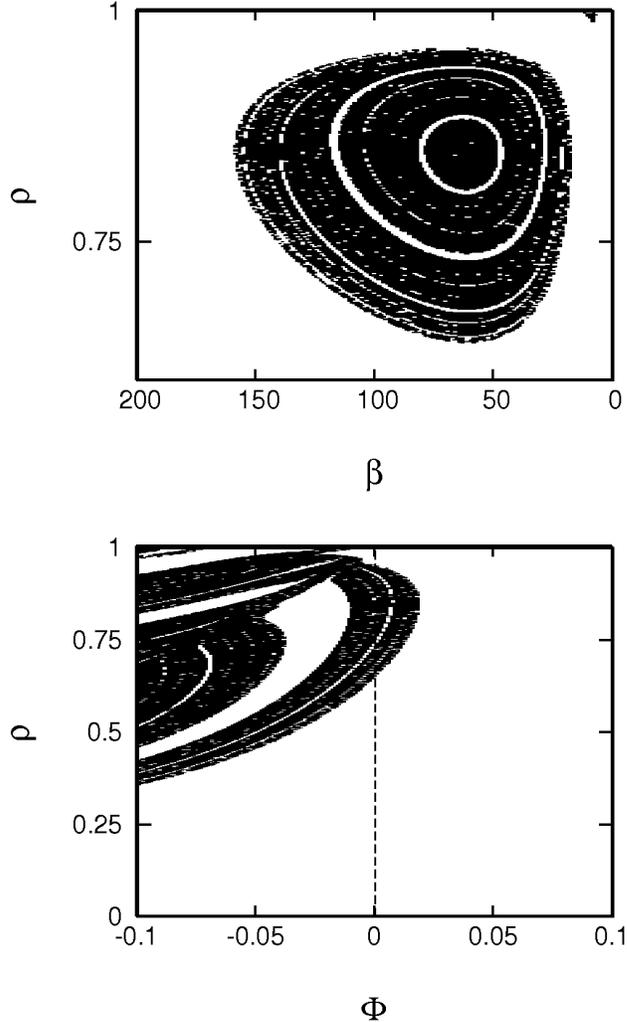,width=15cm, angle=-90}}
\caption{Lyapunov surface. The Lyapunov exponent as a function
of $\protect\beta $ for $\Phi =0.01$ (top), and as a function of $\Phi $ for 
$\protect\beta =25$ (bottom), as obtained from the saddle--point map (%
\protect\ref{mt}). The black regions and curves correspond to a positive
Lyapunov exponent, so that dynamics is then irrregular. The white regions
correspond to a negative Lyapunov exponent associated with regular behavior.
Note a small black, chaotic region for $\protect\rho \lesssim 1$ and low $%
\protect\beta $ in the upper graph.}
\label{figure2n}
\end{figure}

Our model may exhibit a complex dependence on $\rho $ assuming activity
dependent weights. This is expected to occur in many cases, e.g., for
excitable media \cite{exci,julyan}. However, as far as we know, the only
situation with time--dependent connections which is well documented in the
literature concerns the brain. In this case, transmission of information and
computations have repeatedly been reported to be correlated with
activity--induced fast fluctuations of synapses, i.e., our $w_{ij}$'s \cite%
{noise,abb}. For example, it has been observed that the efficacy of synaptic
transmission can undergo short--time increasing (sometimes called \textit{%
facilitation}) \cite{facil,facil2,Wang} or decreasing (\textit{depression}) 
\cite{depre000,9,abb0}, and that these effects depend on the activity of the
presynaptic neuron. Furthermore, it has already been demonstrated that such
processes may importantly affect a network performance \cite%
{cortesNC,facilitation,depre1,bibit,Romani}. Likewise, it seems sensible to
assume that similar short--time variations may occur in other networks
---e.g., reaction--diffusion systems and the cardiac tissue \cite{julyan}---
associated with some efficacy lost after heavy work or with excitations, for
instance.

Motivated by all these facts, and also trying to maintain a well--defined
reference frame, we shall assume that the connection weights are%
\begin{equation}
w_{ij}=\varepsilon _{ij}\overline{w}_{ij}=\varepsilon _{j}\overline{w}_{ij},
\label{c1}
\end{equation}%
where the second equality is introduced for simplicity. Here, $\overline{w}%
_{ij}$ stands for some reference value and $\varepsilon _{j}$ for a random
variable. That is, we are assuming some \textquotedblleft
noise\textquotedblright\ on top of a previous preparation of the connections
designed so that the network can perform some specific function. The
background just described also suggests us to assume that the random
variable in (\ref{c1}) is fluctuating very rapidly so that, on the time
scale for the activity changes, it behaves as stationary with distribution
given, for example, by%
\begin{equation}
p^{\text{st}}\left( \mathbf{s},\varepsilon _{j}\right) =q\delta \left(
\varepsilon _{j}-\Phi \right) +\left( 1-q\right) \delta \left( \varepsilon
_{j}-1\right) .  \label{c2}
\end{equation}%
We shall further assume that $q$ depends on the degree of \textit{order} in
the system at time $t,$ namely, that $q=q\left( \mathbf{\pi }_{t}\right) .$
For the sake of concreteness, our choices here will be that $q\left( \mathbf{%
\pi }\right) =\left( 1+\alpha \right) ^{-1}\sum_{\mu }\pi ^{\mu }\left( 
\mathbf{s}\right) ^{2}$ and that $\overline{w}_{ij}$ is given by the Hebb
prescription. The result is that each node is acted on by an effective field%
\begin{equation}
h_{i}^{\text{eff}}\left( \mathbf{s},\mathbf{w}\right) =\sum_{j\neq i}w_{ij}^{%
\text{eff}}s_{j}  \label{he}
\end{equation}%
with $w_{ij}^{\text{eff}}=\left[ 1-\left( 1-\Phi \right) q\left( \mathbf{\pi 
}\right) \right] \overline{w}_{ij}.$ This amounts, in summary, to assume
short--term variations which change the intensity or capacity of connections
by an amount, either positive or negative, $\Phi $ on the average. More
specifically, one has a decreasing effect for any $\Phi <1$, and enhancement
for $\Phi >1,$ as far as $\Phi >0,$ while $\Phi <0$ induces a change of
sign. For the indicated choices of fields and reference weights, our
framework reduces to the familiar Hopfield--Hebb case for $\Phi =1.$ Note
that it should not be difficult to implement the model for choices other
than (\ref{c1}) and (\ref{c2}).

\section{Description of main results}

Assuming (\ref{he}), it readily ensues from (\ref{mt}) for $M=1$ that $\pi
_{\infty }=F\left( \pi _{\infty };\rho ,\Phi \right) $ and that local
stability requires that $\left\vert \partial F/\partial \pi \right\vert <1,$
where%
\begin{equation}
F\left( \pi ;\rho ,\Phi \right) \equiv \rho \tanh \left\{ \beta \pi \left[
1-\left( 1-\Phi \right) \pi ^{2}\right] \right\} +\left( 1-\rho \right) \pi .
\end{equation}%
Therefore, fixed points are independent of $\rho $ for any $\Phi ,$ but
stability demands that $\rho <\rho _{c}$ with $\rho _{c}=2\left\{ 3\beta \pi
_{\infty }^{2}\left[ \left( {\frac{4}{3}-}\Phi \right) -\left( 1-\Phi
\right) \pi _{\infty }^{2}\right] -\beta +1\right\} ^{-1}.$ The resulting
situation for any $\Phi \neq 1$ is illustrated in Fig. \ref{figure1n}, where
one observes regular behavior, bifurcations and chaotic windows. This
picture cannot occur for fixed weights, e.g., in the Hopfield case. In order
to deepen on the possibility of chaos, we computed the Lyapunov surface from
the analytical solution for $M=1.$ Two sections of this surface are shown in
Fig. \ref{figure2n}. This clearly reveals the existence of chaos above some
degree of synchronization, namely, for $\rho \geq \rho _{\ast }\left( \beta
,\Phi \right) >\rho _{c}\left( \beta ,\Phi \right) $ where the latter marks
the onset of period doubling before irregular behavior. For example, the top
graph shows that, for a small positive value of $\Phi ,$ which corresponds
to some slight depression of connections which occurs more likely the higher
the current system order is, there is a region for large $\beta $
(relatively small temperature, say $T\approx 0.02$ in our arbitrary units)
and $1>\rho \gtrsim 0.8$ for which dynamics may eventually become chaotic.
In the same graph one may notice a tiny chaotic window for $\rho \approx 1$
and $\beta \approx 7;$ this is the case identified previously by us \cite%
{chaos}. The bottom graph, on the other hand, illustrates that chaos is
typically an exception for positive values of $\Phi ;$ it may only occur
then for a rather large fraction of synchronized nodes (large $\rho )$ near $%
\Phi \lesssim 0.$ On the contrary, for negative $\Phi ,$ i.e., when the
order tends to induce changes of sign of the connection intensities, it is
more likely that the system will behave chaotically. It is also to be
remarked that, inside the first, more exterior curve in each graph, there is
a complex pattern of transitions from regular to irregular behavior as one
changes, even very slightly the values of $\rho ,$ $\Phi $ and $\beta .$ As
one may imagine, this situation for very small $M$ gets even more involved
as $M$ increases. Finally, it is noticeable the fact that chaotic switching
among different patterns was recently demonstrated to occur also in the
thermodynamic limit \cite{LT}. The next question is whether such complex
behavior may have some constructive role in natural and man--made networks.
\begin{figure}[ht!]
\centerline{\psfig{file=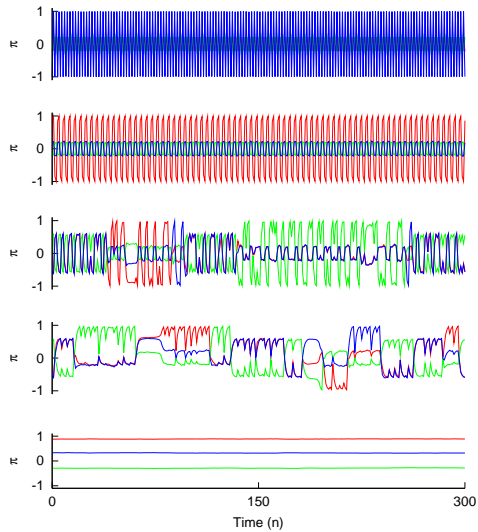,width=6.5cm}}
\caption{Typical Monte Carlo runs. This
shows the overlap as a function of time (in units of $n$ MC trials), during
the stationary regime after equilibration, for $N=1600$ nodes, $\protect%
\beta =50,$ $\Phi =0.004$ and, from bottom to top, $\protect\rho =0,$ 0.60,
0.87, 0.93 and 1.00, respectively. In this case, the onset of period
doubling before irregular behavior is at $\protect\rho =\protect\rho %
_{c}\simeq 0.5.$ This is for $M=3$ correlated patterns (identified here with
different colours). That is, we generated three patterns completely at
random, and then replaced 20\% of the digits in the second and third
patterns with the same number of digits, and flipped digits, respectively,
taken from the first pattern.}
\label{figure3n}
\end{figure}

Different types of behavior the system may exhibit are illustrated by the 
\textit{stationary }Monte Carlo runs in Fig. \ref{figure3n}. This involves
three partially correlated patterns, as explained in the figure caption, and
illustrates, from bottom to top:

\begin{enumerate}
\item For $\rho <\rho _{\ast },$ convergence towards one of the attractors,
namely, fixed points corresponding to the patterns provided. This is
revealed by the fact that one of the overlaps (the red one in this case) is
constantly rather large, close to 1, while the others two are closer to zero
(they differ from zero due to the built correlations between patterns).

\item Irregular behavior with positive Lyapunov exponent for a larger value
of $\rho .$ Notice that changes with time indicate that dynamics is now
unstable and the system activity is visiting the different attractors,
including the negative of some of them or \textit{antipatterns. }

\item A different type of irregular behavior in which, in addition to
visiting different attractors on a large time scale, there are much more
rapid irregular transitions between one pattern and its antipattern.

\item Regular oscillation between one attractor and its negative.

\item Rapid and ordered periodic oscillations between one pattern and its
antipattern when all the nodes are active.
\end{enumerate}

\noindent The cases 2 and 3 are examples of instability--induced switching
phenomena, namely, the system describes heteroclinic paths among the
attractors, and remains different time intervals in the neighborhood of each
of them, as it was previously observed in a related case \cite{chaos}.

\begin{figure}[ht!]
\centerline{\psfig{file=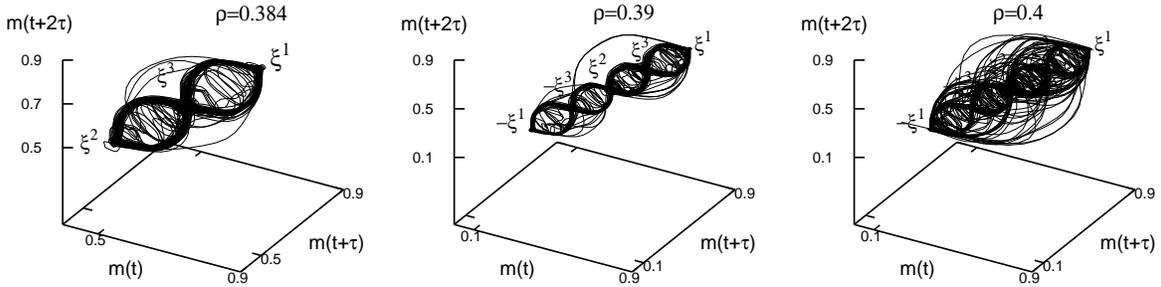, width=15.5cm}}
\caption{Monte Carlo runs that one may interpret as \textit{states of attention} in the network, which illustrates
the possible role of chaos. This shows phase--space trajectories of the mean
firing rate for $N=1600,$ $\protect\beta =167,$ $\Phi =-{\frac{1}{2}},$ and,
from left to right, $\protect\rho =$ 0.384, 0.39 and 0.4. Here, $\protect%
\rho _{c}=0.38,$ and the system stores three patterns, $\protect\xi ^{%
\protect\mu },$ $\protect\mu =1,2$ and 3, as described in the main text.
(These graphs involve a standard false--neighbor method \protect\cite{embed}
with \textit{embedding dimension} $d_{e}=5$ and time delay $\protect\tau=5.) $ }
\label{figure5}
\end{figure}

An interesting fact concerning the nature of the phase space trajectory as $%
\rho $ is varied is illustrated in Fig. \ref{figure5}. This shows time
evolution of the mean firing rate defined as 
\begin{equation}
m=\frac{1}{2N}\sum_{i=1}^{N}\left( 1+s_{i}\right) .
\end{equation}%
Three patterns (and their corresponding antipatterns) are involved here
which consist of a string of 1s, a string with the first 50\% positions set
to 1 and the rest to $-1,$ and a string with only the first 20\% positions
set to 1, respectively. In the course of this Monte Carlo experiment, we
observed that the activity remains wandering around one of the patterns for
any $\rho <\rho _{\ast }$. The choice of pattern depends on the initial
condition. For larger values of $\rho $ within a chaotic window, as for the
three cases shown in Fig. \ref{figure5}, the system tends to visit the other
patterns as well. In particular, the left--most case in the figure ($\rho
=0.384)$ shows visits to the three patterns, and a trajectory which is
structured, namely, there are many jumps between the pairs of more
correlated patterns, and only a few between the most distant ones. Moreover,
the number of jumps between the less correlated patterns tends to increase
as $\rho $ is further increased within the chaotic window. The figure shows
that, for $\rho =0.39$ and 0.40, even the antipatterns are visited; note
that we have that $\mathbf{\xi }^{2}=-\mathbf{\xi }^{2}$. Increasing $\rho $
further, e.g., for $\rho =0.6$ in this specific experiment, the network
surpasses equiprobability of patterns and, eventually, abandons the chaotic
regime to fall into a limit cycle, where it periodically oscillates between
a pattern and its antipattern.

\begin{figure}[ht!]
\centerline{\psfig{file=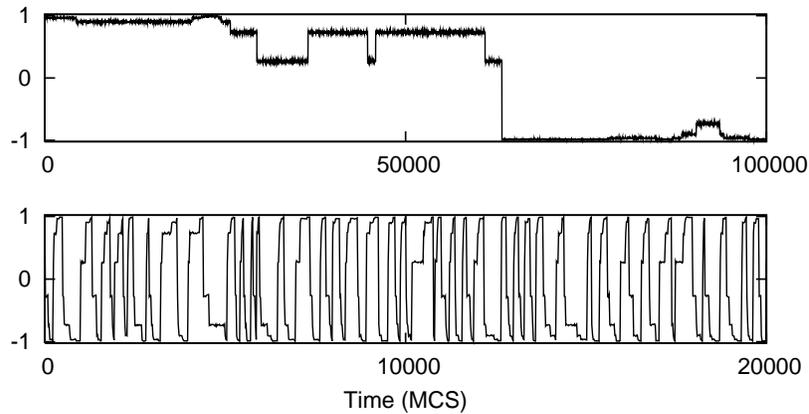,width=11cm}}
\caption{Time series for the overlap $%
\protect\pi $ in the case $\protect\rho =0.632,$ $M=1,$ $\protect\beta %
=\infty $ (zero temperature), $N=3600,$ and $\Phi =$ --0.048 (top) and
--0.065 (bottom) showing chaotic transitions between the associated pattern
and its antipattern. This series correspond to entropies $S\simeq $ 0.37 and
0.9, respectively.}\label{figure7}
\end{figure}

In order to deepen further on the nature of the chaotic switching, we have
computed the normalized power spectra $p\left( \omega \right) $ of the time
series for the mean firing rate $m.$ If one computes the associated entropy 
\cite{bio}, namely, $S=-\sum_{\omega }p_{\omega }\log $.$p_{\omega },$ it
ensues a sharp minimum at $S\simeq 0.37$ for $\Phi =-0.048.$
\begin{figure}[ht!]
\centerline{\psfig{file=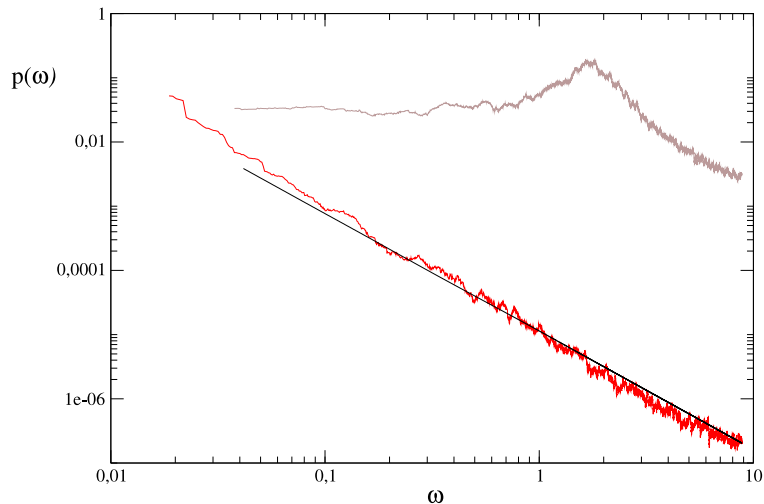,width=10cm}}
\caption{The power spectra corresponding to the two
series in figure \protect\ref{figure7}, i.e., for $\Phi =-0.065$ and $-0.048,
$ respectively, for the upper and lower sets of data. The stright line here
has negative slope 1.9.}\label{figure8}
\end{figure}
The series corresponding to this minimum and, for comparison
purposses, a different one for a much larger entropy are presented in figure %
\ref{figure7}. The power spectra for these two series is presented in figure %
\ref{figure8}. This reveals a qualitative change of behavior, namely, that
(only) the series describing a more efficient chaotic mechanism exhibit a
power law distribution. We are presently analyzing in more detail this
interesting phenomenon. However, we can already illustrate further the
situation as in figure \ref{figure5Dt}.
\begin{figure}[ht!]
\centerline{\psfig{file=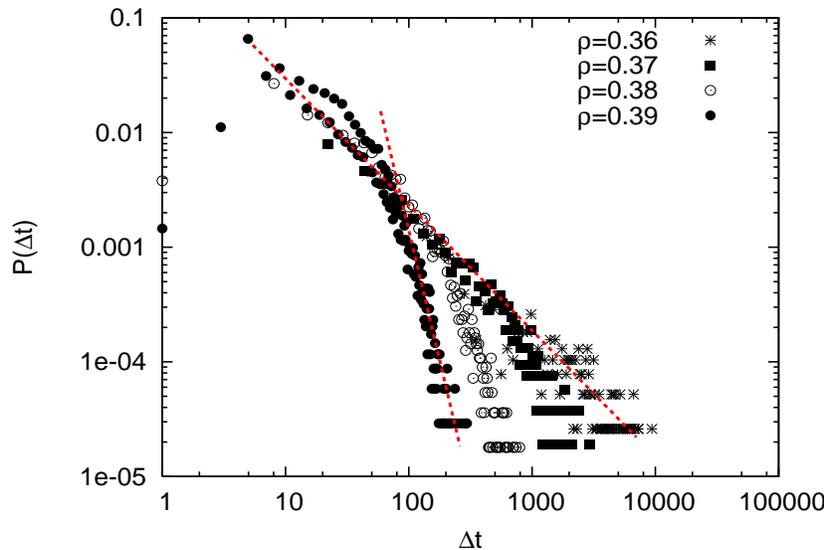, width=11cm}}
\caption{Distribution of the time intervals the network activity stays near
each pattern during the computer experiment described in figura \protect\ref%
{figure5}.}
\label{figure5Dt}
\end{figure}
This shows the distribution for the time intervals the network activity
spends wandering in the neighborhood of a particular attractor.

\section{Discussion}

We have described in this paper details concerning a model network in which
connections are heterogeneously weighted and time--dependent, namely,
correlated to the global activity. As documented above, these two conditions
occur in many natural networks. Furthermore, only a fraction $\rho $ of
nodes are active at each time, so that the rest maintain the previous state.
This occurs in excitable media, for instance.

A main conclusion is that, although the synchronization parameter $\rho $ is
generally irrelevant, varying $\rho $ may greatly modify the system behavior
under certain conditions. The necessary condition is a kind of
susceptibility or sensitivity to external stimuli which greatly favours
dynamic instabilities. It may be achieved in our example by appropriate
tuning of two parameters, $\Phi $ and $\beta .$ The latter is an inverse
temperature which controls the stochasticity of the process. The former
induces either enhancement ($\Phi >1)$ or lowering ($\Phi <1)$ for positive $%
\Phi ,$ or even change of sign for negative $\Phi $, of the intensities of
connections. This process is a very fast one ---as compared with the nodes
changes---, and it occurs more likely the larger the current degree of order
is. The interesting behavior described in this paper washes out if the
connection weights are fixed, even heterogeneously as, for instance, in a
Hopfield--Hebb network, which corresponds here to $\Phi =1.$

Within the most interesting range for its parameters, our model exhibits
heteroclinic trajectories which imply, in particular, a kind of \textit{%
dynamic association}. That is, the network activity either goes to one
attractor for $\rho <\rho _{\ast },$ or else, for larger $\rho ,$ is capable
of an intriguing programme of visits to possible attractors. The dynamic
path followed during these visits may abruptly become chaotic, which seems
the most relevant regime. Besides synchronization of a minimum of nodes,
this requires careful tuning of $\rho ,$ $\beta $ and $\Phi .$ That is, as
suggested by Fig. \ref{figure2n}, there is a complex parameter space which
makes it difficult to predict the ensuing behavior for slight changes of
parameter values. Note in this respect that figure \ref{figure2n} is for $%
M=3 $ patterns only, and that the corresponding picture greatly complicates
as $M $ is increased.

The most interesting behavior of the network consists of \textit{switching}
among attractors. We observe regular switching in some occasions for $\rho
<\rho _{\ast },$ but chaos makes such process much more efficient.
Therefore, our model confirms expectations \cite{caos1,caos2,attent3} that
the instability inherent to chaos facilitates moving to any pattern at any
time, and that chaos and chaotic itinerancy may be the strategy of nature to
solve some difficult problems \cite{caos10,caos11}. Consistent with this, we
have illustrated above a specific mechanism which allows for an efficient
search of the attractors' space. More specifically, we observe a
highly--structured chaotic itinerancy process in which, as illustrated in
Fig. \ref{figure5}, modifying $\rho $ within a chaotic window ---which
requires also tuning $\beta $ and $\Phi $--- one may control the subset of
visited attractors. That is, increasing $\rho $ within the relevant regime
makes the system to visit \textit{more distant} (less correlated)
attractors. In this way the system may perform, for instance, family
discrimination and classification by tuning $\rho $ \cite{algo}. On the
other hand, the complexity of the parameter space for $\rho >\rho _{\ast }$
suggest that one could devise a method to control chaos in these cases. It
is also suggested that one should pay attention to these facts when
determining efficient computational strategies in artificial machines.
Similar switching phenomena, in which the activity describes a heteroclinic
path among saddle states, has already been incorporated in models which thus
simulate experiments on animal olfactory systems \cite%
{Rabi,mazor,huerta,huerta2,attent3,olfato}. Comparable oscillatory activity
has been reported to occur in cultured neural networks \cite{wage} and
ecology models and food webs \cite{hof,van,van2}. This also seems to explain
transitions between atmospheric patterns \cite{crom,stew}, and it is
believed it could account for other natural phenomena as well \cite{attent3}.

Finally, an important feature of the model chaotic itinerancy is illustrated
in figures \ref{figure5Dt} and \ref{figure8}. This reveals the existence of
power--law distributions within the regimes in which the network exhibits
its most interesting behavior. This is the case for the power spectra of
time series and for the time spent in the neighborhood of each attractor for
appropriate values of $\rho .$ This fact suggests that a \textit{critical
condition} which has been called for to explain some of the brain
exceptional behavior \cite{crit1,crit2,crit3,crit4,chialvo} could perhaps
consists of a highly susceptible, unstable and chaotic condition similar to
the one we have described for the model. The ocurrence of power--law
behavior here is consistent with the approaching to zero of associated
Lyapunov exponents, sometimes referred to as \textit{edge of chaos.}

We acknowledge very useful discussions with S. de Franciscis, and financial
support from FEDER--MEC project FIS2005-00791, JA project P06--FQM--01505,
and EPSRC--COLAMN project EP/CO 10841/1.

\end{document}